\documentstyle[epsfig,amsmath,amssymb,graphicx,tabularx]{elsart}
\newcounter{saveeqn}

\def\gsimeq{\,\,\raise0.14em\hbox{$>$}\kern-0.76em\lower0.28em\hbox  
{$\sim$}\,\,}  
\def\lsimeq{\,\,\raise0.14em\hbox{$<$}\kern-0.76em\lower0.28em\hbox  
{$\sim$}\,\,}  
\def\beqy{\begin{eqnarray}}
\def\eeqy{\end{eqnarray}}
\def\bmlet{\begin{mathletters}}
\def\emlet{\end{mathletters}}
\begin{document}
 \begin{frontmatter}  
 
\title{Photodisintegration of Ultra-High-Energy Cosmic Rays revisited}
\author{E. Khan$^1$, S. Goriely$^2$, D. Allard$^1$, E. Parizot$^1$, T.
Suomij\"arvi$^1$}
\author{A. J. Koning$^3$, S. Hilaire$^4$ and M. C. Duijvestijn$^3$}

\address{
$^1$ Institut de Physique Nucl\'eaire, IN$_{2}$P$_{3}$-CNRS/Universit\'e
Paris-Sud, 91406 Orsay, France \\
$^2$Institut d'Astronomie et d'Astrophysique, ULB - CP226, 1050 Brussels,
Belgium \\ 
$^3$Nuclear Research and Consultancy Group, P.O. Box 25, NL-1755 ZG Petten, The
Netherlands \\
$^4$D\'epartement de Physique Th\'eorique et Appliqu\'ee, Service de Physique
Nucl\'eaire, B.P. 12 - F-91680 Bruy\`eres-le-Ch\^atel, France
}

\begin{abstract}
Recent microscopic and phenomenological calculations of giant dipole
resonances for A $\le$ 56 nuclei are presented. The derived
photodisintegration cross sections are exhaustively compared to the
photonuclear data available to date. An accurate description of the data is
found. Our new calculations are also compared with the previous and widely-used
estimates of Puget, Stecker and Bredekamp. The present calculations also include all the possible paths down the nuclear chart. The impact on the photodisintegration of ultra-high-energy cosmic rays (UHECR) is illustrated for a Fe source with typical energies of 10$^{20-21}$eV. At energies around
10$^{20}$eV, the new cross sections are found to modify the UHECR photodisintegration rates. At energies around 10$^{21}$eV, it is recommended to solve a full reaction network to estimate the photodisintegration rate of the UHECR. 
\end{abstract}

\end{frontmatter}

\section{Introduction}

Cosmic-rays have been observed up to energies of $\sim 3\,10^{20}$~eV~\cite{ay94,bi95}, which raises a double problem. First, one has to find an accelerator capable of reaching such ultra high energies. Second, it has been known for almost four decades that the interaction of the cosmic microwave background radiation (CMB) with ultra-high-energy cosmic rays (UHECR) should lead to a sharp decrease in their flux above energies around $10^20$~eV~\cite{gr66,za66}. In the case of protons, which has given rise to the most detailed investigations, this \emph{cut-off} is due to pion photoproduction from CMB photons with energies above the reaction threshold in the proton rest frame. It has been proposed, however, that heavier nuclei could contribute significantly to the high energy part of the spectrum, as originally investigated by Puget, Stecker and Bredekamp (PSB) \cite{pu76,st99}. The propagation of nuclei in the intergalactic medium is also influenced by interactions with the background radiation fields, through photodisintegration reactions.

In the nucleus rest frame, at typical UHECR energies of 10$^{19}$-10$^{21}$
eV, the CMB photons are boosted to energies in the range between a
few hundreds of keV to a few hundreds of MeV. The interaction process
between the UHECRs and the CMB is dominated by the giant dipole resonance
(GDR) at photon energies below 30-50~MeV, and to a lesser extent by the
quasideuteron emission for intermediate energies (between 50 MeV and 150
MeV) and the pion photoproductions at energies above 150 MeV
\cite{pu76,ra96}. In the original PSB model, two major approximations are
performed to estimate the intergalactic UHECR propagation. The first one
concerns the total photoabsorption cross section which is parameterized as a
simple Gaussian function~\cite{pu76,st99}, abruptly cut below the theoretical reaction threshold.
The second one is based on the use of a reduced reaction network, involving only one nucleus for each value of A, to estimate the time evolution of the UHECR composition. More precisely, assuming that the
$\beta$-decay of the unstable nuclei produced by photodisintegration is
always faster than the corresponding photoemission rate, a unique nuclear
path is followed from the initial $^{56}$Fe to the final protons
\cite{pu76}.

The aim of this work is to study the impact of these two approximations on
the properties of the UHECR photodisintegration, and to provide the scientific
community with new accurate determination of the photoreaction rates.
Reaction model developments allow now for an accurate and reliable
systematic estimate of the photodisintegration rate. More specifically,
photoreactions have been extensively studied in the field of
nucleosynthesis, where phenomenological parameterizations of the
photoabsorption cross sections have been optimized during the last decades,
and large-scale microscopic predictions have also emerged \cite{go02,go04}.
New compilations of experimental photoabsorption data also help in defining
the degree of accuracy of the present reaction models to predict the
corresponding cross sections. Important progress has also been performed in
the same field of nucleosynthesis to solve large reaction networks exactly
in order to follow the time evolution of the composition of the material in
given astrophysical sites. Similar tools can therefore be used in the field
of UHECR to test the PSB approximation of a reduced network.

Section 2 describes improved calculations of the photodisintegration cross sections,
using several phenomenological and microscopic descriptions of the dominant $E1$-strength
function. To test the accuracy of such models, they are systematically compared with the
available photonuclear data for nuclei with $A<56$, and with the PSB
Gaussian parameterization. Section 3 is devoted to the study of the impact of the
newly-determined photodisintegration cross section on the propagation of a $^{56}$Fe
source. The approximation of the original PSB reduced network is also tested in Sect. 3 by
comparing the UHECR photodisintegration obtained by solving the full reaction network exactly. 

\section{Calculation of the photodisintegration cross sections}

To describe the changes in abundance of the heavy nuclei as
a result of the interaction of the UHECR with the CMB, a nuclear reaction network including
all interactions of interest must be used.  The chosen set
of nuclear species are coupled by a system of differential equations
corresponding to all the reactions affecting each nucleus, i.e. mainly
photodisintegrations and $\beta$-decays. The rate of change of the number density
$N_{Z,A}$ of nucleus $(Z,A)$ with charge number $Z$ and mass number $A$ can 
be written as
\begin{eqnarray}
\label{eq1}
{dN_{Z,A}  \over dt} &  & =  N_{Z+1,A} \lambda_{\beta}^{Z+1,A} + N_{Z-1,A}
\lambda_{\beta}^{Z-1,A} \cr 
& &+ N_{Z,A+1} \lambda_{\gamma,n}^{Z,A+1} + N_{Z+1,A+1} \lambda_{\gamma,p}^{Z+1,A+1} 
+ N_{Z+2,A+4} \lambda_{\gamma,\alpha}^{Z+2,A+4}  \cr 
& &+ N_{Z,A+2} \lambda_{\gamma,2n}^{Z,A+2}  + N_{Z+2,A+2} \lambda_{\gamma,2p}^{Z+2,A+2} 
+ N_{Z+4,A+8} \lambda_{\gamma,2\alpha}^{Z+4,A+8}  \cr 
& &+ N_{Z+1,A+2} \lambda_{\gamma,np}^{Z+1,A+2}  + N_{Z+2,A+5}
\lambda_{\gamma,n\alpha}^{Z+2,A+5}  +  N_{Z+3,A+5}
\lambda_{\gamma,p\alpha}^{Z+3,A+5}  \cr  
& & - N_{Z,A} \left\lbrack\lambda_{\beta}^{Z,A}
+ \sum_x
  \lambda_{\gamma,x}^{Z,A} \right\rbrack~, 
\end{eqnarray}

\noindent where $\lambda_{\beta}^{Z,A}$ is the $\beta$-decay rate of nucleus $(Z,A)$ and
$\lambda_{\gamma,x}^{Z,A}$ its photoerosion rate followed by the
emission of a single neutron ($x=n$), proton ($x=p$) or  $\alpha$-particle ($x=\alpha$) or
 the emission of multiple particles such as $2n$, $2p$, $2\alpha$, $np$, $\dots$,
including all open channels for a given photon energy distribution.

The CMB photon density n($\epsilon$) depends only on the UHECR Lorentz
factor $\gamma=E/Mc^{2}$ (where E is the UHECR energy, and M its mass)
\cite{st99}. The calculations of the CMB density as a function of the photon
energy $\epsilon$ in the nucleus rest frame show that photon energies
overlap with the nuclear GDR for $\gamma$ ranging from 5.10$^9$ to
10$^{12}$. In the nucleus rest frame, the photodisintegration rate
$\lambda_{\gamma,x}$ can be estimated from the cross section
$\sigma_{\gamma,x}(\epsilon)$ by

\begin{equation}
\lambda_{\gamma,x}=\int{n(\epsilon)~\sigma_{\gamma,x}(\epsilon)~c~d\epsilon}~
\label{eq2}
\end{equation}
\noindent where $c$ is the light speed.

All nuclei lighter than the seed nuclei and located between the valley of
stability and the proton drip line must be included in the network. Under
the most natural astrophysical assumptions, UHECRs are accelerated out of
the ambient gas, possibly enriched in Fe close to neutron stars or depleted
in metals (i.e. nuclei heavier than H) if significant photodisintegration
occurs during the acceleration stage itself. Therefore, if nuclei are indeed
present among the UHECRs, it is expected that they typically include the
most abundant elements found in the interstellar medium, i.e. essentially
lighter than Fe. The interaction of UHECRs with the CMB is thus expected to
include all possible nuclei resulting from the photodisintegration of the
heaviest species and therefore involve all stable and neutron-deficient
unstable isotopes with $A \lsimeq 56$.

One of the most complete compilation of photonuclear data is provided by the
2000 IAEA atlas (2000) \cite{ia00}. Nevertheless, as far as elements of
interest in the propagation of UHECR are concerned, only a limited set of
photonuclear cross sections are known, namely the total photoabsorption
cross section as a function of energy for about 10 nuclei and the integrated
total photoabsorption cross section for no more than 16 nuclei. All the
remaining rates must therefore be estimated on the basis of theoretical
reaction models.

\subsection{The $E1$-strength function}

The uncertainties involved in any cross section calculation are not so much related to the
model of formation and de-excitation of the compound nucleus itself, than to the
evaluation of the nuclear quantities necessary for the calculation of the transmission
coefficients. The total photon transmission coefficient characterizing the probability to
excite by photoabsorption a compound nucleus excited state is obviously one of the key
ingredients for the evaluation of the photoreaction rates. In the specific
astrophysical conditions considered here, i.e for UHECR energies of $10^{19-21}$~eV, this
function is  dominated by the $E1$ transition which is classically estimated within the
Lorentzian representation of the GDR. Experimental photoabsorption data confirm the
simple semi-classical prediction of a Lorentzian shape at energies around the resonance
energy $E_{GDR}$. One the most widely used form of the
$E1$-strength function is described by the Brink-Axel Lorentzian model
\cite{br55,ax62}
\begin{equation}
T_{E1}(\varepsilon_{\gamma})= \frac{8}{3} \frac{NZ}{A} \frac{e^2}{\hbar c}
\frac{1+\chi}{mc^2} ~\frac{\Gamma_{\mathrm{GDR}}
\,\varepsilon_{\gamma}^4}
{(\varepsilon_{\gamma}^2-E_{\mathrm{GDR}}^2)^2+\Gamma_{\mathrm{GDR}}^2 
\,\varepsilon_{\gamma}^2},
\label{eq_tg}
\end{equation}
%
where $E_{GDR}$ and $\Gamma_{GDR}$ are the energy and width of the GDR, $m$ is
the nucleon mass and $\chi \simeq 0.2$ is an exchange-force contribution to the dipole sum
rule.
 
The Lorentzian description is known to be less satisfactory at energies away from the GDR
peak, and in particular fails to describe the low-energy experimental data, namely the
radiation widths and gamma-ray spectra \cite{mc81,ko90}. Various improvements have been
brought to the Lorentzian form, mainly by including an energy-dependence of the GDR width
capable of modifying the low-energy behavior of the
$E1$-strength \cite{go98,mc81,ko87,ko90}. For this reason, the photon transmission
coefficient is  most frequently described in the framework of the phenomenological
Kopecky-Uhl generalized Lorentzian model \cite{ko90}. In this approximation, the
GDR width of Eq.~(\ref{eq_tg}) is replaced by an energy-dependent width of the form 
$\Gamma(\varepsilon_{\gamma})=\Gamma_{GDR} [\varepsilon_{\gamma}^2 +4 \pi
T^2]/E_{GDR}^2$, where $T$ is the nuclear temperature and equals zero in the case of
photoabsorption reactions. This model is the most widely used for practical
applications, and more specifically when global predictions are requested for large sets
of nuclei. It also requires the determination of the GDR peak energy and width to be
predicted from some underlying model for each nucleus. For practical applications, these
properties are either taken directly from experimental compilations (e.g
\cite{ia00,ripl}) whenever available, or obtained from a droplet-type model
\cite{my77} or some experimental systematics \cite{ripl}. 

The phenomenological Lorentzian approach suffers, however, from shortcomings of various
sorts, and most particularly its lack of reliability when dealing with exotic
nuclei or energies away from the GDR peak.  For this reason, models
of the microscopic type have been developed which are hoped to provide a reasonable
reliability and predictive power for the
$E1$-strength function. Attempts in this direction have been conducted within models like
the thermodynamic pole approach \cite{ripl}, the theory of finite Fermi systems or the
Quasiparticle Random Phase Approximation (QRPA) \cite{ka83}. The spherical
QRPA model making use of a realistic Skyrme interaction has even been used recently for
the large-scale derivation of the $E1$-strength function.
In such models, mean field calculations such as Hartree-Fock BCS
(HFBCS)\cite{kh00} and Hartree-Fock Bogoliubov (HFB)
\cite{gr01} are performed in order to describe the nucleus ground state.
On top of these calculations, the QRPA is used
to describe in a microscopic way the GDR. The linear response
theory allows to predict the dipole strength of the excited nucleus. The only input of
these mean-field models is the nucleon-nucleon interaction, including the Skyrme part,
and the pairing component. Details can be found in \cite{kh02}. 

Global HFBCS+QRPA and HFB+QRPA calculations for practical applications were developed in
\cite{go02,go04} and shown to predict the location of the GDR in close agreement with
experimental data and also to reproduce satisfactorily the average resonance capture data
at low energies. These aforementioned QRPA calculations have been performed for all the
$8\le Z\le 110$ nuclei lying between the two drip lines. Note that the latest HFB+QRPA
calculation is using the BSk7 Skyrme nucleon-nucleon interaction, initially derived to
reproduce at best the measured masses \cite{go03}.

\subsection{The photodisintegration cross sections}

The photoreaction cross sections are estimated with the code named Talys
\cite{ko04} which takes into account all types of direct, pre-equilibrium and compound
mechanisms to estimate the total reaction probability as well as the
competition between the various open channels. The photoreaction cross
section is estimated at energies up to 50 MeV. The calculation includes
single particle (nucleons and alpha) as well as multi-particle emissions.
All the experimental information on nuclear masses, deformation and
low-lying states spectra is considered, whenever available. If not, global
nuclear level formul\ae, and nucleon and alpha-particle optical model
potentials are considered to estimate the particle transmission coefficients
and the nuclear level. Details on the codes and the nuclear physics input
(ground state properties, nuclear level densities, optical potential) can be
found in the above mentioned references
\cite{ko03,ko04}. These various nuclear inputs are known to affect the
photodisintegration rates mainly around the corresponding threshold
energies, but have a relatively lower impact than the $\gamma$-ray strength
function. To estimate the accuracy of the different approaches available for
the evaluation of an $E1$-strength function, the four models presented
in the previous sub-section are considered, namely the Lorentzian
\cite{ax62}, the generalized Lorentzian \cite{ko90}, the HFBCS+QRPA
\cite{go02} and the HFB+QRPA
\cite{go04}. In addition to the $E1$ component, the $E2$,
$M1$ and $M2$ contributions are also included in the calculation of the
$\gamma$-ray strength as prescribed by \cite{ripl}, but are not varied due
to their lower impact in comparison with the GDR. The quasideuteron process
is also neglected in the present work due to the limited photon energy
range \cite{ra96}.

\subsection{Comparison with the photonuclear data}

The photoreaction cross sections estimated as explained above are now
compared with available experimental data \cite{ia00} for nuclei with
A$\le$56. It should be recalled that, even for stable nuclei, the data on
such nuclei are scarce: a cross section measurement,
such as ($\gamma$,1nx) or ($\gamma$,abs), is available  at various energies for less than
half of the stable nuclei with 12$\le$A$\le$56. Total photoabsorption cross sections
around the GDR peak energy are available for 10 nuclei, while the integrated
total photoabsorption cross section is known for 16 nuclei \cite{ia00}. The
exhaustive comparison performed to compare the predictions with
the data is illustrated below.

\begin{figure}
	\centering \includegraphics[width=9cm]{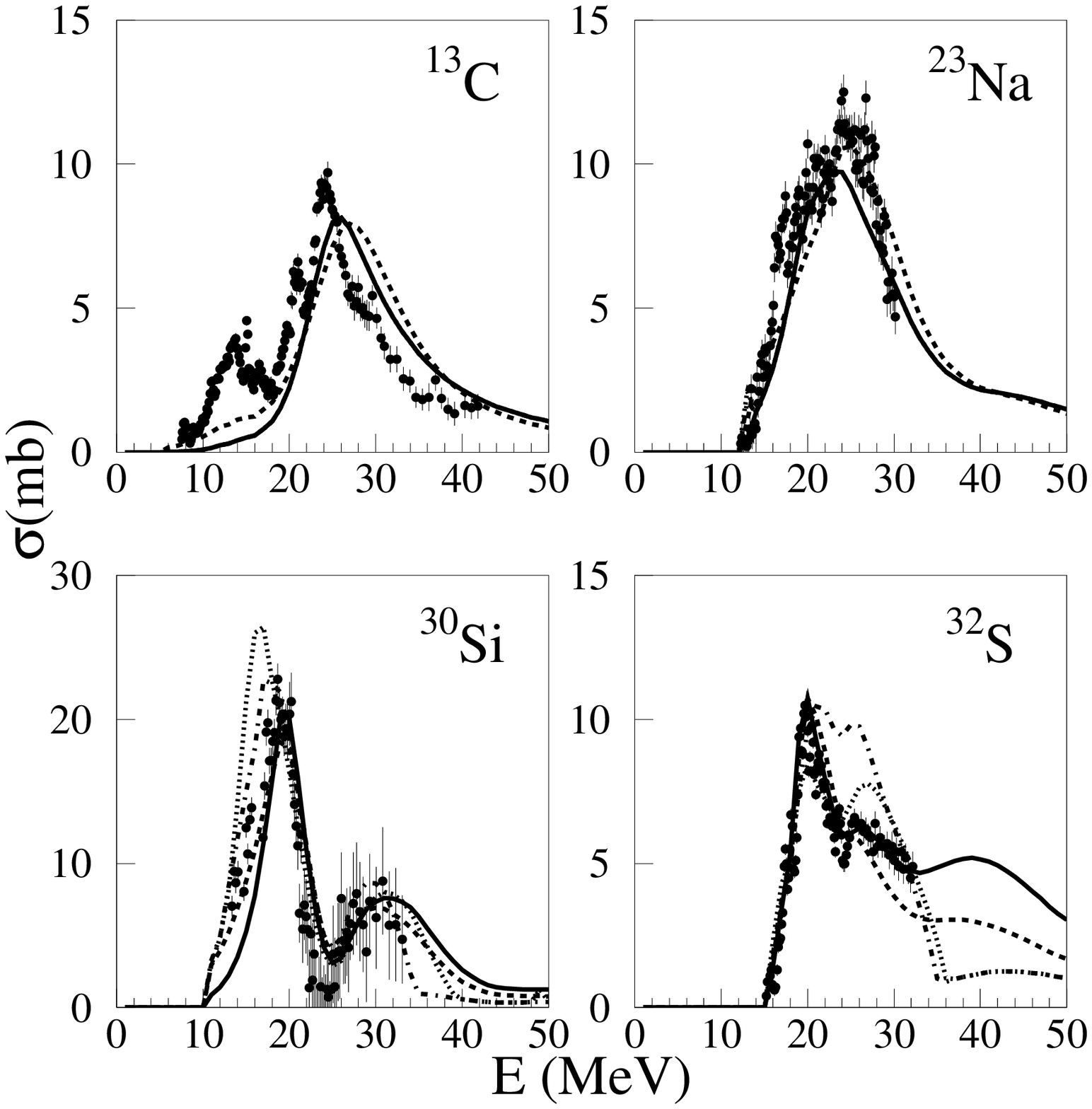}
\caption{Measured photoabsorption cross sections ($\gamma$,1nx) state,
compared to the predictions of the four models: Lorentzian (dashed line),
generalized Lorentzian (solid line), microscopic HFBCS+QRPA (dotted line)
and microscopic HFB+QRPA (dash-dot line)}
\label{fig:cs}
    \end{figure}

\begin{figure}
	\centering \includegraphics[width=9cm]{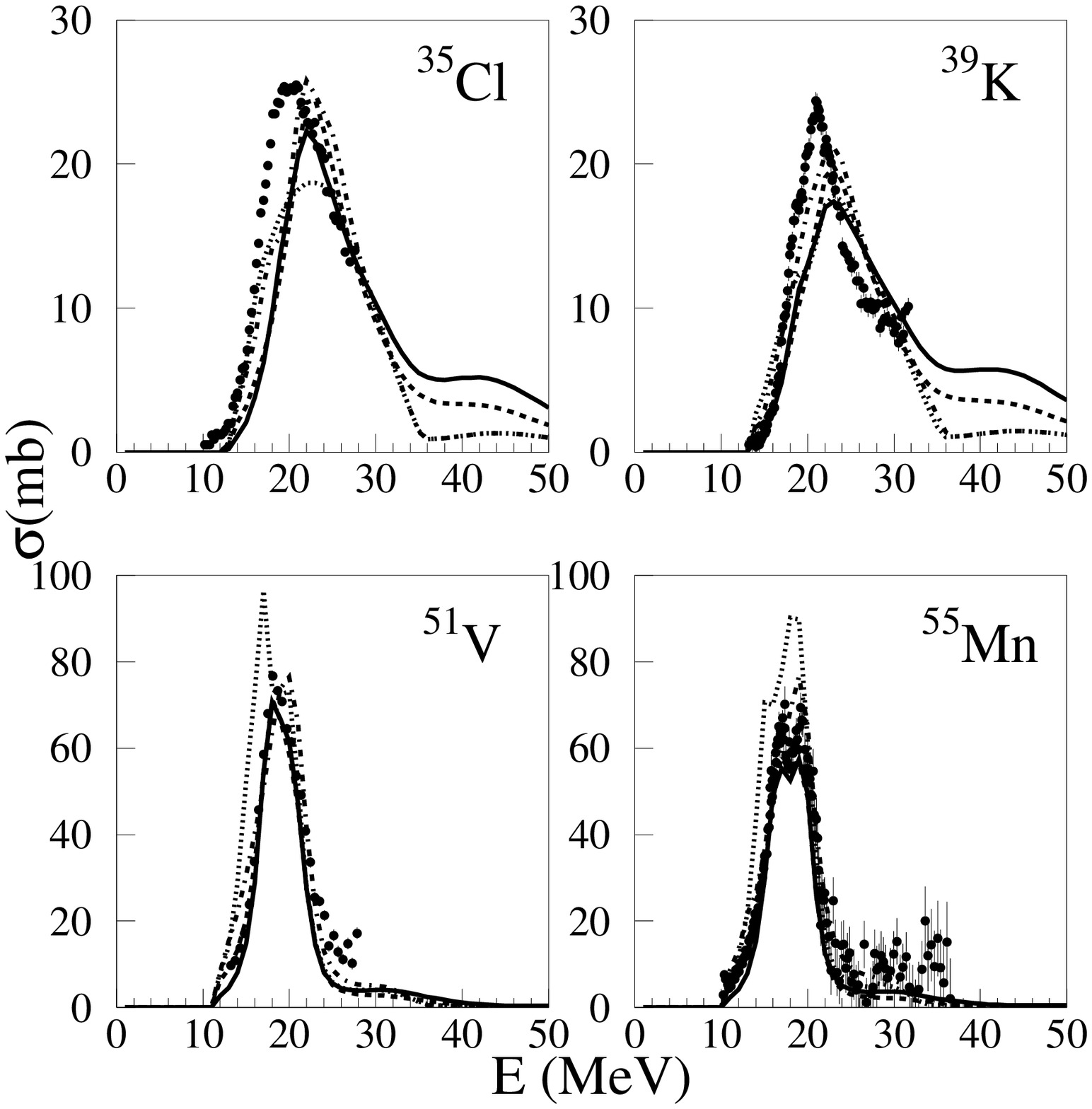}
\caption{Measured photoabsorption cross sections ($\gamma$,1nx) state,
compared to the predictions of the four models: Lorentzian (dashed line),
generalized Lorentzian (solid line), microscopic HFBCS+QRPA (dotted line)
and microscopic HFB+QRPA (dash-dot line)}
\label{fig:csb}
    \end{figure}

To test the model predictions, we first consider the ($\gamma$,1nx) cross
sections, i.e the photodisintegration leading to a single neutron emission,
but possibly also to other extra ejectiles. This set is of particular
importance for our present study since the one-nucleon emission is the
dominant process for photon energies in the 0-50 MeV range. The model
predictions based on four different prescriptions for the $E1$-strength
function is compared in Figs.~\ref{fig:cs} and ~\ref{fig:csb} with
experimental data for 8 nuclei, namely $^{13}$C, $^{23}$Na, $^{30}$Si,
$^{32}$S, $^{35}$Cl, $^{39}$K, $^{51}$V and $^{55}$Mn. Globally, the
calculations are found to describe the data accurately, from light to heavy nuclei, even though a few isotopes still suffer from imperfect description, such as $^{13}$C. In most cases, however, the behaviour of the cross-section is remarkably well described, being confined within the experimental error bars over the whole energy range. For astrophysical applications, this
degree of accuracy is quite satisfactory, especially in view of the
remaining uncertainties pertaining to the origin of UHECRs (distribution of
sources, injection spectra, initial composition) and to their propagation in
extragalactic space, depending on the essentially unknown strength and
topology of the magnetic effects.

In some cases, the QRPA prediction is found to overpredict the data,
especially at high energies. This is the case for the HFB+QRPA prediction
of the $^{32}$S($\gamma$,1nx) reaction, or for the HFBCS+QRPA
results for $^{30}$Si, $^{51}$V and $^{55}$Mn. On average, both
phenomenological Lorentzian approaches reproduce better the experimental
data. In the case of $^{13}$C and $^{23}$Na, both the microscopic
calculations and the Lorentzian-based predictions agree with each other. It
should however be recalled here that the Lorentzian formulas make use of the
experimental total photoabsorption peak cross section, peak energy and width for the cases
presented here, while the QRPA models do not.

\begin{figure}
	\centering \includegraphics[width=9cm]{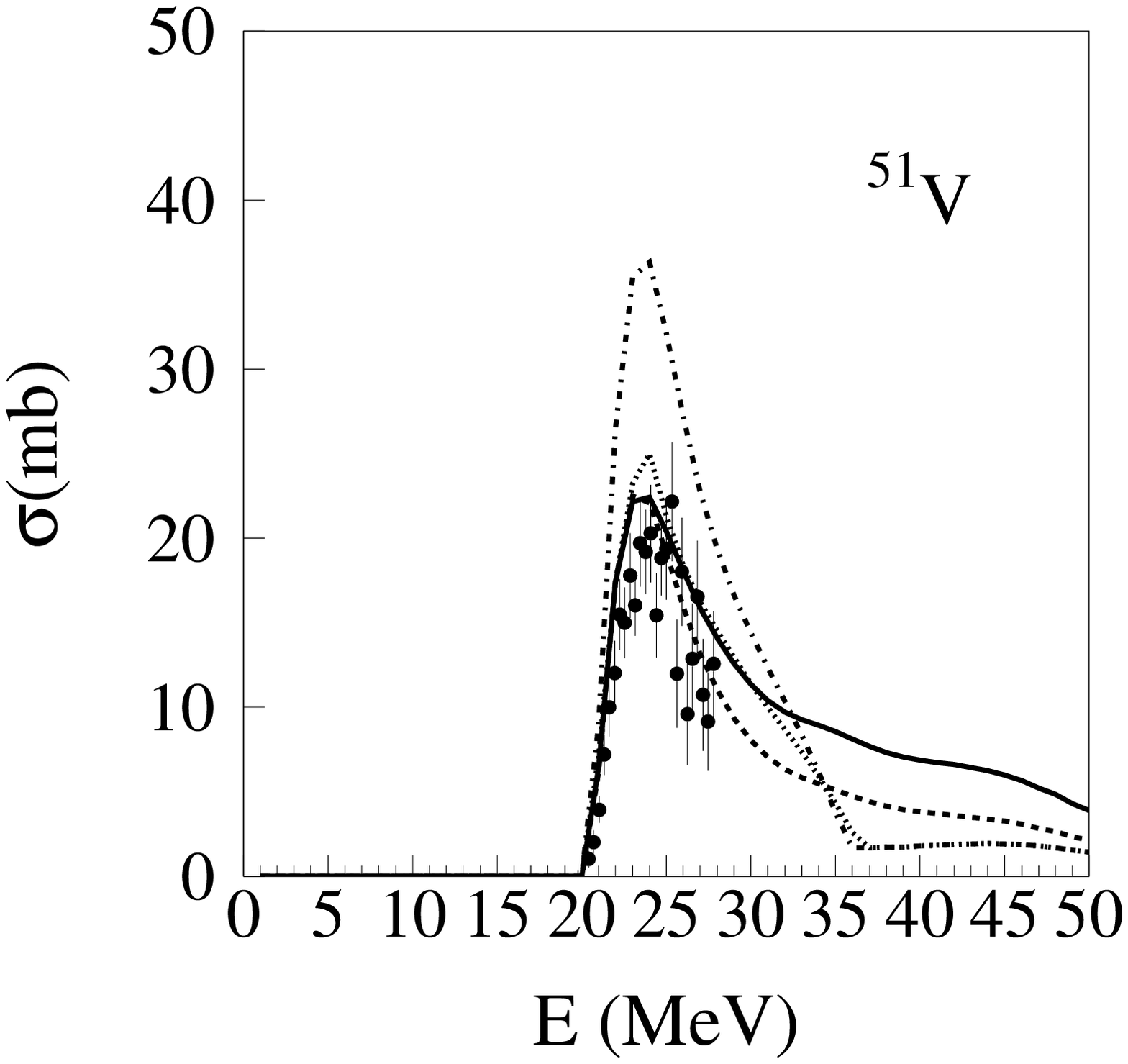}
\caption{Measured photoabsorption cross sections
$^{51}$V($\gamma$,2nx)compared to the predictions of the four models:
Lorentzian (dashed line), generalized Lorentzian (solid line), microscopic
HFBCS+QRPA (dotted line) and microscopic HFB+QRPA (dash-dot line)}
\label{fig:cs2}
    \end{figure}

In order to illustrate the accuracy of multi-nucleon photoemission
predictions, the $^{51}$V($\gamma$,2nx) measured cross section is compared
to the four calculations in Fig. \ref{fig:cs2}. The agreement is good and
all models predict similar cross sections, except the HFB+QRPA model which
overestimates the cross section. The errors on such measurement are
relatively large and prevent one from considering this data as a strong constraint on
the models.

The ($\gamma$,p) cross sections provide complementary insight on the
accuracy of the models. The integrated cross section is available for about
15 nuclei of interest \cite{ia00}. The description of the data is found to be
reasonable, and not to differ from one model to another significantly.
An average agreement within a factor of 2 is obtained on the integrated cross section.

In view of the present comparison between the model predictions and the available
experimental data, the generalized Lorentzian model is adopted for further calculations of
the UHECR propagation. Since the nuclei of interest for the propagation of UHECR are located
close to the valley of stability (see Sect.~3.2), experimental data provide a relatively
accurate systematic determination of the GDR properties and no significant deviation from
the phenomenological parameterization of the Lorentzian type is expected.

\subsection{Comparison with the PSB parameterization}

\begin{figure}
	\centering \includegraphics[width=9cm]{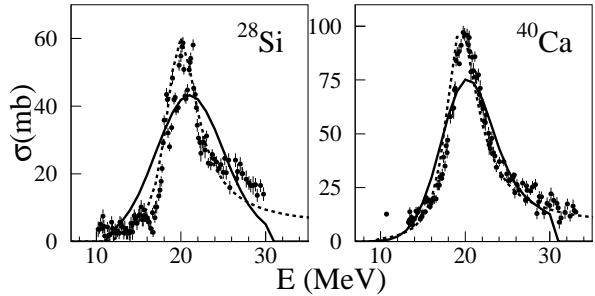}
\caption{Measured total photoabsorption cross sections, compared to the predictions of the PSB
models (solid line) and the generalized Lorentzian (dashed line)}
\label{fig:cs3}
    \end{figure}

The generalized Lorentzian cross sections are now compared with the PSB
Gaussian prescription. This comparison is made for total photoabsorption data. Although
($\gamma$,abs) cross sections are available for a limited number of nuclei (e.g. Ahrens et
al. \cite{ah75}), it allows a direct comparison of the $\gamma$-ray strength,
regardless of the other nuclear inputs in the reaction calculation. The 
$^{28}$Si and $^{40}$Ca photoabsorption cross sections are compared in Fig.~\ref{fig:cs3}.

In their original work \cite{pu76}, PSB calculate the total photoabsorption
cross section, relying on the total integrated cross section, available for
about one third of the nuclei of interest in the UHECR problem. This
modelling of the GDR was performed up to an energy of 30 MeV.  Though not
strongly different from the Lorentzian calculations, it is clear that the
PSB agreement with experimental data is not as good as the one obtained with
the Lorentzian parameterization. This may be explained by two features of
the PSB modelling. First the PSB parameterization assumes a Gaussian shape
for the GDR, which can differ significantly from the
experimentally-confirmed Lorentzian shape. Second, the width of the GDR is
frequently overestimated by the PSB parameterization, compared to the data.
An exhaustive comparison between the PSB and the Lorentzian parameterization
for the $\sim$ 50 nuclei used in the PSB model, shows also the general trend
of a greater width (typically 2 MeV) for the PSB parameterization. The large
PSB width can lead to significant discrepancies, especially close to the
one-neutron separation energy threshold located around 8 MeV. The GDR
contribution close to the neutron threshold plays an important role since it
triggers the photodisintegration of UHECRs having just enough energy to be
affected by the CMB photons while propagating throughout the universe. An
accurate description of this threshold is thus necessary to derive the exact shape
of the resulting turnover in the UHECR propagated spectrum.

The larger width also leads to a larger integrated value of the cross
section between 0 and 30 MeV than in the Lorentzian case. On the other hand
the total integrated cross section between 0 and 50 MeV is rather similar in
both models since the Lorentzian prescription extends further up to 50 MeV
while the PSB Gaussian prescription is limited to energies below 30~MeV.
Note that both models predict a similar location of the peak energy since
they consider experimental information or experimentally-based systematics.

\section{Impact on the UHECR photodisintegration}

The intergalactic UHECR photodisintegration is calculated on the basis of the
rates (Eq.~\ref{eq2}) derived from the Talys cross
sections described in Sect.~2.  All stable and neutron-deficient
unstable nuclei with A$\le$56 are included in the reaction network
(Eq.~\ref{eq1}). The photodisintegration of A$\le$4 nuclei are not
considered. Only the interaction with the CMB is
considered, the influence of the infrared background radiation being
negligible in most cases of interest
\cite{st99}. For illustrative purposes, we restrict ourselves to the case of
a UHECR source made of $^{56}$Fe only. Other astrophysically relevant cases will be discussed elsewhere \cite{al05}. Note that a full propagation model for UHECRs (including pair production, statistical fluctuations, etc.) is not the purpose of this paper. Most recent calculations can be found, e.g., in~\cite{an98,ep98}.

\subsection{Impact of the photodisintegration cross sections}

\begin{figure}
\centering \includegraphics[width=15cm]{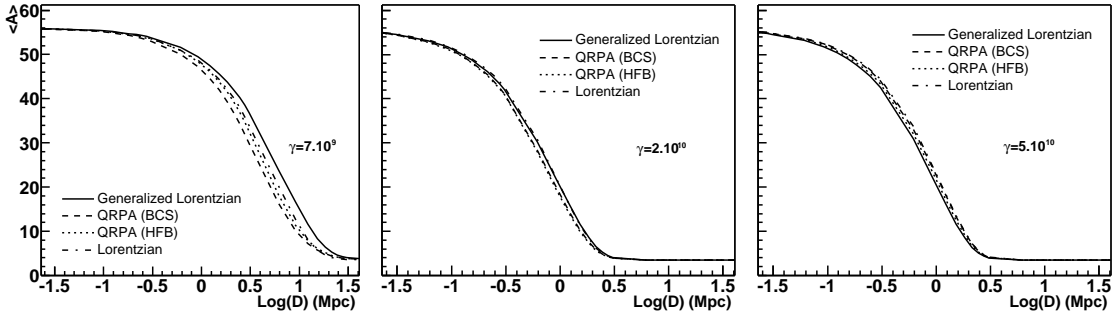}
\caption{Evolution of the average mass number $<A>$ with
respect to the distance of a $^{56}$Fe source for three different Lorentz
factors. Left: $\gamma=7\,10^9$ ($E=3.6\times 10^{20}$ eV); center:
$\gamma=2\,10^{10}$ ($E=10^{21}$eV); right: $\gamma=5\,10^{10}$
($E=2.61\times 10^{21}$eV).}
\label{fig:prop}
    \end{figure}

Fig. \ref{fig:prop} shows the evolution of the average mass number $<A>$ as
a function of the distance from the $^{56}$Fe source, calculated with the
four GDR prescriptions described in Sect.~2. For a given source distance,
$<A>$ is the average value of the calculated nuclei abundances : 

\begin{equation}
<A>=\frac{\sum_{i}N_{i}A_i}{\sum_{i}{N_{i}}}
\end{equation}
where $N_i$ is the number density given by Eq. (\ref{eq1}). The full
reaction network~(Eq.~\ref{eq1}) is solved at each time step, taking into
account all the open photoemission channels, i.e ($\gamma$,n), ($\gamma$,p),
($\gamma$,$\alpha$), ($\gamma$,2n), ($\gamma$,2p), ($\gamma$,2$\alpha$),
($\gamma$,np), ($\gamma$,n$\alpha$), ($\gamma$,p$\alpha$). In other words,
the abundance of each type of nucleus is derived by taking into account the
contribution of all the production channels, from the source nucleus
downwards the table of nuclides, with the appropriated weight derived
according to the corresponding cross sections. The values obtained are thus
equilibrium values, representing the composition which would result from the
propagation of an infinite number of nuclei up to the time considered. For
smaller UHECR samples, of course, Poissonian fluctuations of the numbers of
each nuclear species are expected.

Three illustrative values of the Lorentz factors $\gamma$ are considered.
The value $\gamma=7\,10^9$ corresponds to an initial energy of the $^{56}$Fe
nucleus of $E=3.6\times 10^{20}$ eV. In this regime, only the lowest energy
part of the $E1$-strength overlaps with the photon density n($\epsilon$).
The photodisintegration rate is therefore sensitive to the position of the
photoemission threshold located at the neutron separation energy ($S_n\sim$
8 MeV). The value $\gamma=2\,10^{10}$ corresponds to a total energy of
$E=10^{21}$eV. The photon density is maximum around the neutron separation
energy, and the photodisintegration rate is sensitive to the location of the
GDR peak energy. Finally, the large value of $\gamma=5\,10^{10}$
($E=2.61\times 10^{21}$eV) corresponds to a regime mainly sensitive to the
energy-integrated photodisintegration cross section.

Fig. \ref{fig:prop} shows that iron nuclei with $\gamma=7\,10^{9}$ propagate
up to distances 10 times larger than in the higher energy cases, before
being stripped from all their nucleons. This is due to the small overlap
between the photon density and the photodisintegration cross section in the
low-$\gamma$ case. In this regime, the distance of propagation is sensitive
to the low-energy tail of the $E1$-strength function and the use of
different prescriptions leads to significant differences in the propagation
distance. In contrast, the high-$\gamma$ cases mainly depend on the GDR peak
location or integrated photoabsorption, and for this reason the propagation
distance is less sensitive to the photoreaction details. In addition, all
calculations predict similar integrated photoabsorption cross sections since
they all make use of the energy-weighted sum rule to normalize the dipole
strength \cite{rs80}. The spread observed between the different curves in
Fig.~\ref{fig:prop} reflects the impact of the remaining nuclear
uncertainties on the UHECR propagation.

It should be noted that during propagation, the Lorentz factor of each
fragment of the original nucleus remains essentially the same, since in a
first approximation the total energy is evenly distributed among all
nucleons. The whole propagation thus occurs at constant $\gamma$ and the
interaction regime (at the threshold, around the peak or through the whole
range of the cross section) remains the same along the fragmentation
process. It should be reminded, however, that additional energy losses occur
in real astrophysical situations, due to e$^{+}$e$^{-}$ pair production as
well as pion production. These important refinements are not included here
because we focus on the specific influence of the nuclear cross sections,
but they will be studied in detail in a forthcoming paper.

\begin{figure}
	\centering \includegraphics[width=15cm]{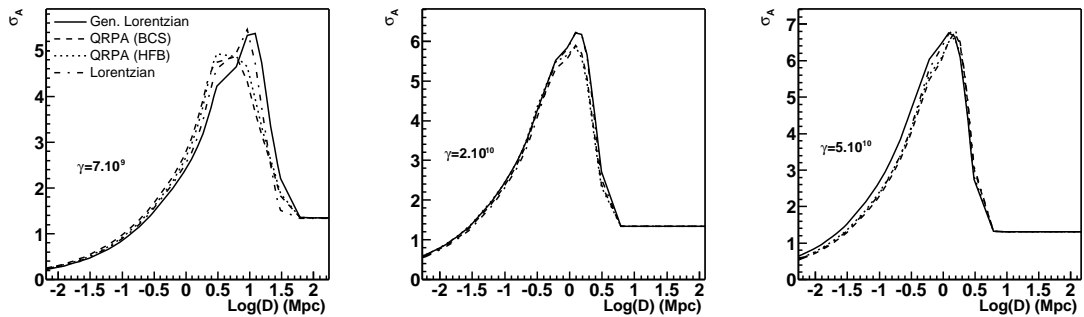}
\caption{Evolution of the standard deviation
of the A distribution with respect to the distance of a $^{56}$Fe source for
three different Lorentz factors. Left: $\gamma = 7\,10^9$
($E=3.6\times 10^{20}$eV); center: $\gamma=2\,10^{10}$ ($E=1.0\times 10^{21}$eV);
right: $\gamma=5\,10^{10}$ ($E=2.61\times 10^{21}$eV).}
\label{fig:prop2}
    \end{figure}

The UHECR mass distributions are displayed in Fig.~\ref{fig:prop2} and show
interesting features. The standard mass deviation is defined as :
\begin{equation}
\sigma_{A}=\sqrt{\frac{\sum_{i}N_{i}(A_i-<A>)^{2}}{\sum_{i}{N_{i}}}}
\end{equation}

All the $E1$-strength parameterizations are seen to lead to similar
predictions for the shape and magnitude of the standard deviation from the
average mass number $<A>$ as a function of the distance to the source. This
overall mass distribution mainly depends on the relative competition between
the various open photoemission channels, rather than on the absolute
photoabsorption rate. This conclusion is valid irrespective of the energy of
the source. At maximum spread, the average deviation from the mean value
$<A>$ reaches values as large as 7 mass units, which means that many
different nuclei are found with comparable abundances as secondaries of the
parent $^{56}$Fe nucleus. In the specific case of $\gamma=7\,10^9$, the
mass distribution ranges from Ne to Fe isotopes at a distance $D=2.5$~Mpc.

\subsection{Photodisintegration path}

\begin{figure}
	\centering \includegraphics[width=15cm]{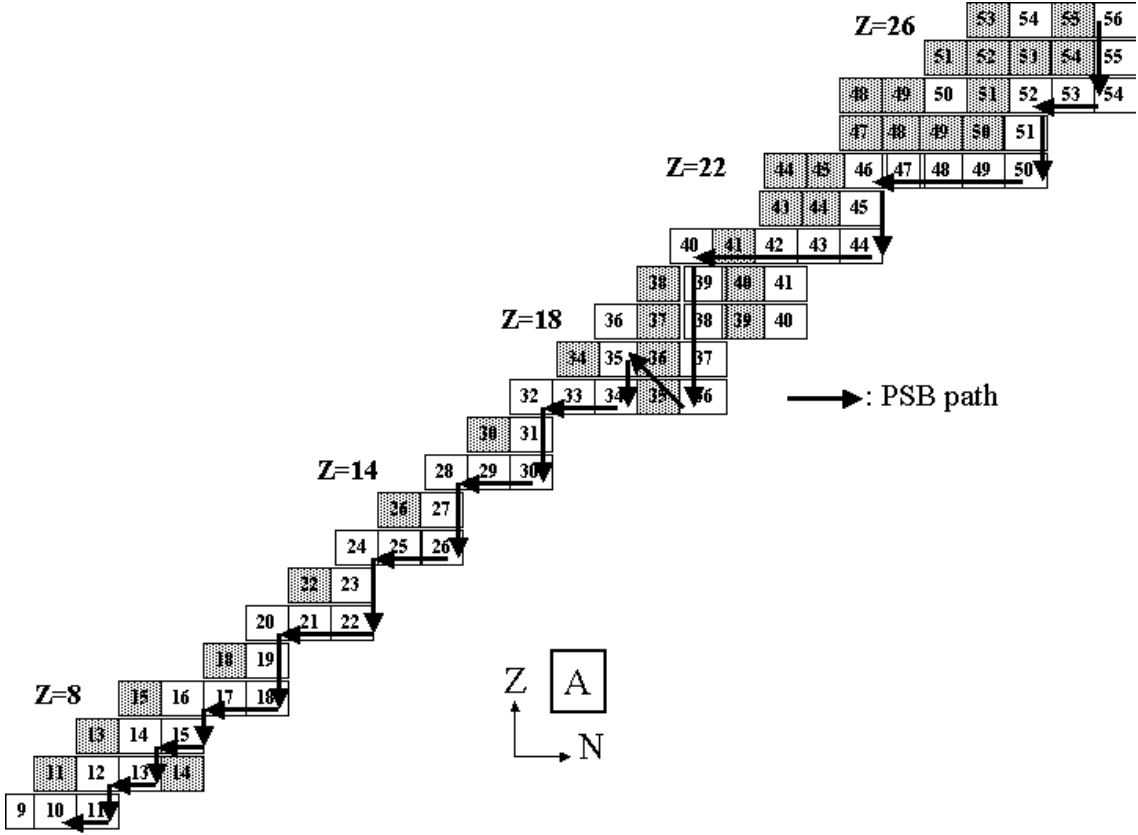}
\caption{Nuclei involved in the
photodisintegration process of $^{56}$Fe nuclei for $\gamma$=2.10$^{10}$.
Unstable nuclei are in shaded squares, and the PSB path is indicated by the
arrows. The mass number of each nucleus is written in the corresponding
square.}
\label{fig:chart}
    \end{figure}

In Sect. 3.1, the UHECR propagation distance has been estimated making use
of the full reaction network (Eq.~\ref{eq1}). Previous calculations were
based on the reduced PSB path illustrated in Fig.~\ref{fig:chart}. In this
approximation, only one stable isotope is considered per isobaric chain and
the corresponding isobars (i.e. nuclei with the same A) are not affected by competitive channels. However,
as shown in Fig.~\ref{fig:chart}, about 85 nuclei are involved in the
$^{56}$Fe photodisintegration at $\gamma$=2.10$^{10}$ and numerous open
channels including $\beta$-decay can compete (the Lorentz dilation of time
allows $\beta$-unstable nuclei with half-lives of the order of the hour to survive over a Mpc scale, and thus have a chance to interact with a CMB photon). Most of the stable nuclei involved in the
photodisintegration process have more neutrons than protons. Neutron
emissions are therefore favoured and the corresponding unstable nuclei will
$\beta^+$-decay towards the valley of stability. Note that we consider here
that a given nucleus is involved in the reaction network if its calculated
abundance amounts about 10\% of the most produced one at any given time
during the photodisintegration process.

Significant differences can therefore be expected between our new calculation
(Fig.~\ref{fig:prop}) and the original PSB results based on the reduced path and the
Gaussian parameterization of GDR strengths. In particular, as seen in Fig.~\ref{fig:chart},
for heavy nuclei (A$\ge$45), about 70\% of the nuclei are shortcut by the simplified
PSB path. For light nuclei (A$\le$45), less than 42\% of the nuclei are bypassed.

\begin{figure}
	\centering \includegraphics[width=15cm]{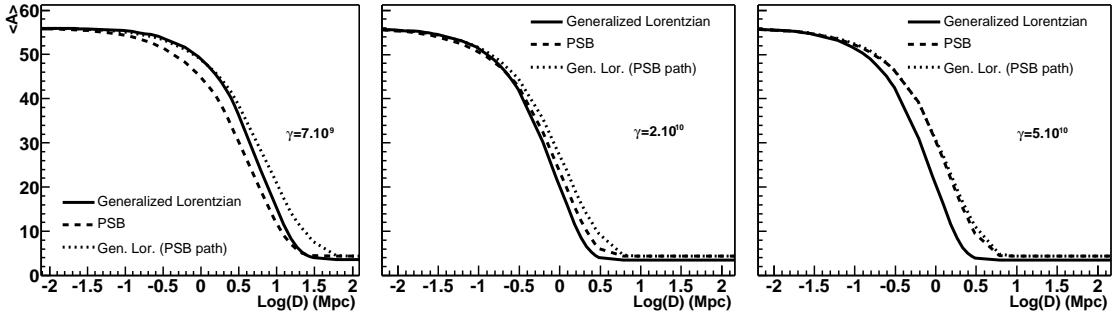}
\caption{Evolution of the average A with
respect to the distance of a $^{56}$Fe source for three different Lorentz
factor. Left : $\gamma=7\,10^9$ ($E= 3.6\times 10^{20}$ eV); center :
$\gamma=2\,10^{10}$ ($E= 1.0\times 10^{21}$eV); right : $\gamma=5\,10^{10}$ ($E=
2.61\times 10^{21}$eV).}
\label{fig:prop4}
    \end{figure}

Fig.~\ref{fig:prop4} displays the average mass numbers $<A>$ with respect to
the source distance for the three $\gamma$ regimes. For each regime, three
calculations are shown, namely (i) the full network calculation based on the
generalized Lorentzian rates (solid line), (ii) the reduced PSB path
calculation with the same generalized Lorentzian rates (dotted line), and
(iii) the original PSB results based on the PSB path and Gaussian
photodisintegration rates (dashed lines). The effect associated with the
type of reaction network adopted is seen to be significant and to increase
slightly with the UHECR energy. The simplified PSB path leads to a longer
propagation distance with respect to the full network calculation,
independently of the nuclear input considered. This path effect is stronger
for heavier than for lighter nuclei. For A $\le$ 45, the curves show similar
slopes. For the heavy species, the major differences stem from the large
number of nuclei excluded from the PSB path, while the full reaction network
calculations show that many isobars contribute to the nuclear flow. Within
an isobaric chain, the photodisintegration cross section is usually larger
for high Z-values, so that nuclei on the PSB path propagate up to large
distances.

The differences between the dashed and dotted lines in Fig.~\ref{fig:prop4} reflect the
impact of the newly-determined photoreaction rates with respect to the widely used PSB
rates. This comparison also confirms the previous conclusion that the cross section effect
is attenuated at high energies due to similar integrated photoabsorption cross sections.
However, the effect of the low-energy $E1$-strength around the threshold against particle
emission remains significant, as seen in the low-$\gamma$ case. Both the path and the
cross section have an impact on the propagation distance, whereas at high
$\gamma$ values, the path effect is the only one to remain.

\section{Conclusions}

Four phenomenological and microscopic models have been used to predict the
photodisintegration rates for nuclei with A $\le$ 56 nuclei on the basis of
the reaction code called Talys. The four models reproduce accurately the
photonuclear data, namely the one-particle emission, two-particle emission,
and total photoabsorption data. The original PSB description of the
total photoabsorption cross section often overestimates the GDR width,
implying a larger GDR contribution at low energies with respect to
experimental data.

We studied in particular the case of a $^{56}$Fe source and found that the new photoabsorption cross sections mainly modify the disintegration of UHECRs at relatively low Lorentz factors ($\sim 7\,10^9$), which is relevant to the shape of the cosmic-ray energy spectrum around the GZK suppression. The key ingredient proves to be a precise description of the GDR threshold. For higher Lorentz factors, details of the nuclear photoreaction rates do not impact much on the propagation distance. In contrast, it is found that solving a full reaction network can modify the results obtained with simpler nuclear paths, most particularly for the A $\ge$ 45 nuclei.  Large spreads in the mass distribution reaching $\pm$ 7 mass units are obtained when use is made of the full reaction network.

The photodisintegration cross sections calculated with the generalized Lorentzian model
are tabulated and made available to the scientific community via the nuclear astrophysics
library at {\it http://www-astro.ulb.ac.be}. These include the ($\gamma$,n),
($\gamma$,p), ($\gamma$,$\alpha$), ($\gamma$,2n), ($\gamma$,2p), ($\gamma$,2$\alpha$),
($\gamma$,np), ($\gamma$,n$\alpha$), ($\gamma$,p$\alpha$) \ldots for all nuclei with
$12\le A\le 56$.

The new approach to UHECR photodisintegration described in this paper will be applied to sensible astroparticle situations in forthcoming works, both using astrophysically motivated source compositions and taking into account the intergalactic magnetic fields.

\smallskip
\noindent{\bf Acknowledgments} S.G. is FNRS Research Associate.
This work has been performed within the scientific collaboration (Tournesol)
between the Wallonie--Bruxelles Community and France. \\

\end{document}